\documentclass[9pt,conference]{IEEEtran}

\usepackage{premeables}

\usepackage{amsmath}
\usepackage{amssymb}
\usepackage{pifont}
\newcommand{\cmark}{\text{\ding{51}}}%
\newcommand{\xmark}{\text{\ding{55}}}%

\usetikzlibrary{quantikz2}
\tikzset{global scale/.style={
    scale=#1,
    every node/.style={scale=#1}
  }
}
\usepackage{tabularx}
\usepackage{makecell}
\usepackage{multirow}

\usepackage{mathrsfs} 
\usepackage{listings}
\usepackage{algorithm}
\usepackage[noend]{algpseudocode}

\definecolor{mygray}{rgb}{0.8,0.8,0.8}

\usepackage{enumitem}
\setlist{parsep = 0.3\parskip,
         itemsep = 0pt plus 1pt,
         topsep = 0pt plus 1pt,
         }
         
\usepackage[T1]{fontenc} 

\definecolor{codegreen}{rgb}{0,0.6,0}
\definecolor{codegray}{rgb}{0.5,0.5,0.5}
\definecolor{codepurple}{rgb}{0.58,0,0.82}
\definecolor{backcolour}{rgb}{0.95,0.95,0.92}

\lstdefinestyle{mystyle}{
    backgroundcolor=\color{backcolour},   
    commentstyle=\color{codegreen},
    basicstyle=\ttfamily\footnotesize,
}

\newcommand{\code}[1]{\texttt{\footnotesize{#1}}}
\usepackage{hhline}

\newif\ifcomment
\commenttrue

\ifcomment
    \newcommand{\hongxiang}[1]{\textcolor{blue}{[Hongxiang: #1]}}
\else
\newcommand{\hongxiang}[1]{}
\fi

\newcommand{\figref}[1]{Fig.~\ref{#1}}
\newcommand{\tabref}[1]{Table~\ref{#1}}

\title{Versatile Cross-platform Compilation Toolchain for\\Schr\"odinger-style Quantum Circuit Simulation}

\makeatletter
\newcommand{\linebreakand}{%
  \end{@IEEEauthorhalign}
  \hfill\mbox{}\par
  \mbox{}\hfill\begin{@IEEEauthorhalign}
}
\makeatother

\author{\IEEEauthorblockN{Yuncheng Lu, Shuang Liang, Hongxiang Fan, Ce Guo, Wayne Luk, Paul H. J. Kelly} 
\IEEEauthorblockA{Department of Computing, Imperial College London, London, UK\\ 
Email: \{yuncheng.lu19, shuang.liang, hongxiang.fan, c.guo, w.luk, p.kelly\}@imperial.ac.uk} }

\newcommand{\CAST}{\texttt{CAST}}

\begin{document}

\maketitle

\ifdefined\hpcacameraready 
  \thispagestyle{camerareadyfirstpage}
  \pagestyle{empty}
\else
  \thispagestyle{plain}
  \pagestyle{plain}
\fi

\newcommand{\hpcaheight}{0mm}
\ifdefined\eaopen
\renewcommand{\hpcaheight}{12mm}
\fi

\begin{abstract}

While existing quantum hardware resources have limited availability and reliability, there is a growing demand for exploring and verifying quantum algorithms. Efficient classical simulators for high-performance quantum simulation are critical to meeting this demand.
However, due to the vastly varied characteristics of classical hardware, implementing hardware-specific optimizations for different hardware platforms is challenging.

To address such needs, we propose CAST (Cross-platform Adaptive Schr\"odiner-style Simulation Toolchain), a novel compilation toolchain with cross-platform (CPU and Nvidia GPU) optimization and high-performance backend supports.
CAST exploits a novel sparsity-aware gate fusion algorithm that automatically selects the best fusion strategy and backend configuration for targeted hardware platforms. CAST also aims to offer versatile and high-performance backend for different hardware platforms. To this end, CAST provides an LLVM IR-based vectorization optimization for various CPU architectures and instruction sets, as well as a PTX-based code generator for Nvidia GPU support.

We benchmark CAST against IBM Qiskit, Google QSimCirq, Nvidia cuQuantum backend, and other high-performance simulators. On various 32-qubit CPU-based benchmarks, CAST is able to achieve up to 8.03x speedup than Qiskit. On various 30-qubit GPU-based benchmarks, CAST is able to achieve up to 39.3x speedup than Nvidia cuQuantum backend.

\end{abstract}

\section{Introduction}

The theory of quantum physics has long been hypothesized as the foundation for a computational tool~\cite{feynman1982simulating, deutsch1985quantum}, offering exponential advantages over classical computers for various algorithms including integer factorization~\cite{shor1994algorithms}, optimization problem solving~\cite{farhi2014quantum, tilly2022variational}, and quantum or hybrid machine learning~\cite{biamonte2017quantum, schuld2019quantum}. 
The potential of quantum computing has driven the development of quantum processor prototypes by both industry and research institutes, such as IBM~\cite{kim2023evidence} and D-Wave~\cite{king2023quantum}.
However, existing quantum computers usually suffer from limited system size, restricted connectivity, and significant noise. 
These issues hinder their reliable deployment for executing deep circuits required by many algorithms, such as Amplitude Estimation~\cite{suzuki2020amplitude} and Hamiltonian Evolution Simulation~\cite{low2017optimal}.

Meanwhile, significant research efforts have focused on developing new quantum algorithms under the assumption of reliable quantum computation~\cite{tian2023recent, blekos2024review}.
Nevertheless,
verifying these algorithms and demonstrating their performance at scale is challenging due to the aforementioned limitations.
Therefore, there is a growing demand for simulators based on classical computers as alternatives to quantum hardware.
High-performance and reliable classical simulators play a vital role in identifying the potential of quantum computers, particularly in scenarios where quantum computers claim supremacy in tackling problems that are intractable for classical systems~\cite{boixo2018characterizing, arute2019quantum, liu21closing, bremner2017achieving}. 
The performance of simulations also becomes crucial for various applications. For instance, Variational Quantum Algorithms (VQA)~\cite{cerezo2021variational} require optimizing parametrized quantum circuits, which often implies thousands of runs on the same circuit structure.
Similarly, Hamiltonian Evolution Simulation relies on the decomposition of the time-evolution operator~\cite{hatano2005finding}, which induces a significant circuit depth.

However, simulating the quantum world using classical computers has been recognized as a challenging problem. To perfectly mimic the physics of the quantum world, it requires either exponentially increasing memory or time with respect to system size~\cite{aaronson2016complexity}.
Among current simulation approaches, Schrödinger style simulations are often used to facilitate the comprehensive verification and evaluation of quantum algorithms.
These methods require exponential memory to keep track of all quantum state information and are capable of handling all kinds of quantum operations.


Although various Schrödinger style simulators have been introduced by both industry~\cite{qiskit2024, cirq2020, pennylane2018} and academia~\cite{jones2019quest, efthymiou2021qibo, suzuki2021Qulacs}, they are suffering two key issues in obtaining accessible and high-performant simulations. 
First, the absence of advanced and versatile simulation-oriented gate rewriting and fusion optimization hampers their scalability for large quantum circuits.
Second, existing simulation tools do not offer comprehensive cross-platform backend support, and there is an urgent need for an easy-to-use cross-platform solution to accommodate the diverse needs of users with varying hardware resources.
To address these challenges, we propose an adaptive and cross-platform (with CPU and Nvidia GPU support) compilation toolchain for high-performance and scalable quantum circuit simulation named \CAST{} (Cross-platform Adaptive Schr\"odinger-style Simulation Toolchain).
\CAST{} introduces a custom intermediate layer and a cost-model based adaptive gate fusion algorithm. The algorithm automatically selects the best fusion configuration according to the characteristics of targeted CPU and GPU platforms. 
\CAST{} incorporates high-performance kernel generators that automate backend support by emitting vectorized LLVM IR codes for a versatile SIMD solution on CPU platforms and PTX codes for utilizing CUDA cores.

Overall, this work makes the following contributions:

\begin{itemize}
    \item \textbf{A novel adaptive gate fusion technique:} This technique considers the sparsity patterns and categories with an agglomerative scheme to apply gate fusion. We also adopt a cost-model based optimization that automatically selects fusion configuration based on targeted hardware. As a result, our tool can apply CPU or GPU-specialized gate fusion optimization to best match the characteristics of the target hardware.
    \item \textbf{A dynamic kernel generation approach:} \CAST{} enables customized optimization to reduce the number of operations needed to perform quantum circuit simulation. Together with the sparsity-aware fusion technique, \CAST{} achieves fast and energy-efficient quantum circuit simulation.
    \item \textbf{A cross-platform quantum simulation framework:} \CAST{} serves as a cross-platform automated compilation toolchain for high-performance quantum simulation on CPU and Nvidia GPU platforms, featuring adaptive front-end optimization algorithms and high-performance backend support.
\end{itemize}


\begin{table*}[tb]
    \caption{Comparison of various classical quantum circuit simulators.}
    \label{tab:149}
    \begin{minipage}{\linewidth}
    \renewcommand{\arraystretch}{1.1}
    \centering
    \begin{tabular}{c|c|c|c|c|c|c|c}
      & \multirow{2}{*}{\makecell{Precision}} 
      & \multirow{2}{*}{CPU Backend SIMD Support\footnote{``External'' means this backend depends on external pre-compiled library, such as numpy's C++ backend.}} 
      & \multirow{2}{*}{\makecell{CUDA\\Support}}
      & \multicolumn{3}{c|}{Gate Fusion}
      & \multirow{2}{*}{\makecell{Dynamic\\Kernel}} 
      \\\cline{5-7}
      & & & & Size & Sparsity & Parametrized & \\\hline\hline
    QuEST~\cite{jones2019quest} & \code{f32} and \code{f64} & \xmark{} & Builtin & \xmark{} & \xmark{} & \xmark{} & \xmark{} \\\hline
    Qibo~\cite{efthymiou2021qibo} & \code{f32} and \code{f64} & External & CuPy & \cmark{} & Partial\footnote{Qibo provides \code{qibojit} as a just-in-time engine.} & Limited & \xmark{} \\\hline
    Qulacs~\cite{suzuki2021Qulacs} & \code{f32} and \code{f64} & \code{f64} AVX2 only & Builtin & \cmark{} & \xmark{} & \cmark{} & \xmark{} \\\hline
    Qiskit~\cite{qiskit2024} & \code{f32} and \code{f64} & \code{f64} AVX2 only & cuQuantum & \cmark{} & Partial\footnote{Qiskit's gate fusion algorithm does not consider general gate matrix sparsity, but it separates diagonal gates and general unitary gates.} & \xmark{} & \xmark{} \\\hline
    QPanda~\cite{qpanda2022} & \code{f64} only & \code{f64} AVX2 only & cuQuantum & \cmark{} & \xmark{} & Limited & \xmark{} \\\hline
    Cirq~\cite{cirq2020}\footnote{Cirq and QSimCirq (sometimes QSim) are both developed by Google. While Cirq has its own simulation backend, QSimCirq provides a high-performance backend specifically optimized for SSE, AVX2, and AVX512-capable hardware for simulating Cirq circuits.}  & \code{f32} and \code{f64} & External & \xmark{} & \xmark{} & \xmark{} & \xmark{} & \xmark{} \\\hline
    QSimCirq~\cite{cirq2020} & \code{f32} only & \code{f32} AVX2 and \code{f32} AVX512 & cuQuantum & Up to 6-qubit & \xmark{} & Limited & \xmark{} \\\hline
    CUDA Quantum~\cite{kim2023cuda} & \code{f32} and \code{f64} & External & cuQuantum & Up to 6-qubit & \xmark{} & Limited & \xmark{} \\\hline
    CAST (ours) & \code{f32} and \code{f64} & \cmark{}\footnote{Tested to support NEON, SSE, AVX2, and AVX512, both \code{f32} and \code{f64} precision.} & Builtin & \cmark{} & \cmark{} & \cmark{} & \cmark{}
    \end{tabular}
\end{minipage}

\end{table*}




\section{Background and Motivation}\label{sec:background}
\subsection{Statevector-based Simulation}
Statevector-based simulation keeps track of all \emph{amplitudes} in the statevector. An amplitude is a complex number, and an $n$-qubit statevector is a length-$2^n$ complex vector with norm 1. 
Let $\psi$ be the length-$2^n$ complex vector that represents an $n$-qubit statevector. Applying a single-qubit gate \gU{} on qubit $k$ is equivalently performing the matrix-vector multiplication specified by 
\begin{equation}
    \big(\gI^{\otimes(n-k-1)}\otimes\gU{}\otimes \gI^{\otimes{k}}\big)\psi.
\end{equation}
This expression represents multiplying a $2^n\times 2^n$ matrix with a $2^n\times 1$ vector. As the matrix is highly sparse, we can equivalently perform $2^{n-1}$ matrix-vector multiplications in sequential, each time between $\gU{}$ and two of the amplitudes of $\psi$ to improve the efficiency. That is,
\begin{equation}\label{eq:330}
    \gU{}\mat[b]{\psi_{\alpha(t)}\\\psi_{\beta(t)}},\quad t=0,\cdots,2^{n-1}-1.
\end{equation}

More generally, if $\gU$ is a $k$-qubit gate represented by a $2^k\times 2^k$ unitary matrix, then applying $\gU$ on an $n$-qubit statevector $\psi$ consists of $2^{n-k}$ matrix-vector multiplications
\begin{equation}
    \gU\mat[b]{\psi_{\alpha_0}(t)\\\vdots\\\psi_{\alpha_{2^k-1}(t)}},\quad t = 0,\cdots,2^{n-k}-1.
\end{equation}




\subsection{Ahead-of-time Kernel Preparation}
While a dense single-qubit gate can have up to 8 non-zero scalars, each Pauli gate only has two. Simulators with support to only general quantum gates inevitably lose optimizations for Pauli gates. Therefore, for better performance, a traditional approach is to hand-code popular gate kernels ahead of time. Some simulators such as QSimCirq even write vector intrinsics to gain more performance on selected CPU platforms.

However, such a kernel preparation scheme induces extreme workload in writing different kernels to support every gate. With gate fusion, the workload is even greater. Fused gates usually act on more than two qubits, so their sparsity patterns are much more diverse and unpredictable. 
The pattern of loading and storing vector amplitudes also grows exponentially with respect to the gate size. To reduce workload, some simulators supporting gate fusion, such as Qulacs, QPanda, and QSimCirq, invoke general multi-qubit kernel on fused gates. Some other ones, such as Qiskit, separate diagonal matrices by providing additional diagonal multi-qubit gate kernels. However, none of them could fully appreciate the sparsity patterns of fused gates.

\subsection{Motivation}

\begin{figure}
\centering
\includegraphics[width=0.9\linewidth]{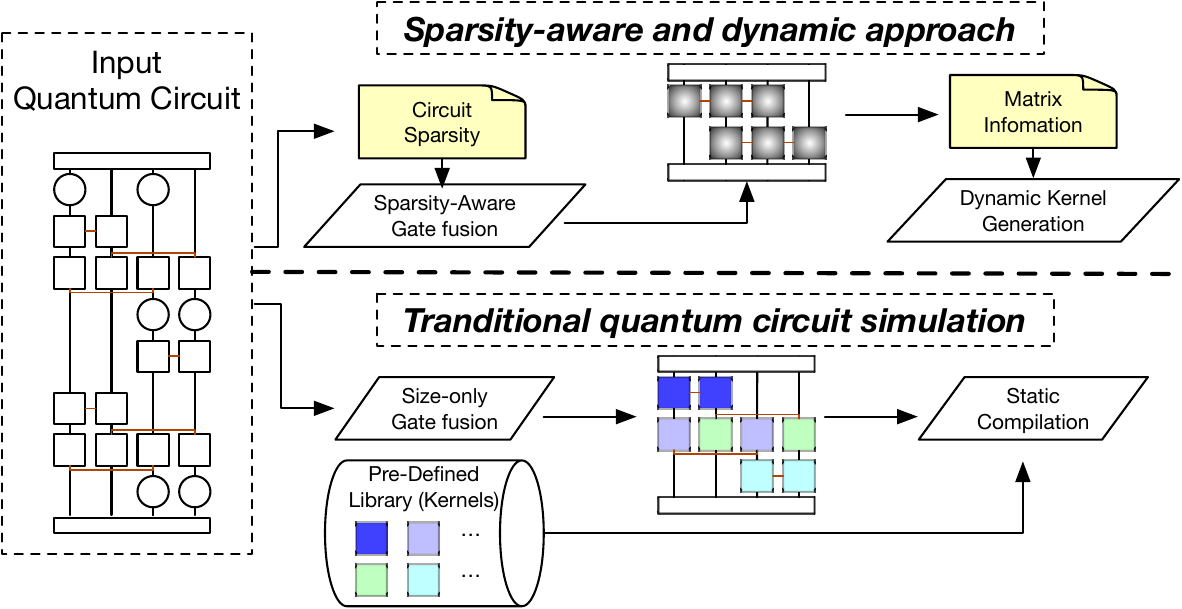}
\vspace{-0.8em}
\caption{The proposed CAST approach exploits sparsity information and runtime-matrix information to improve simulation performance.}
\label{fig:motivation}
\end{figure}

To facilitate the quantum simulation on classical computers,
various quantum circuit simulators have been developed by both industry and academia~\cite{jones2019quest, efthymiou2021qibo, suzuki2021Qulacs, qiskit2024, qpanda2022, cirq2020}. However, as summarized in~\tabref{tab:149}, most of these tools overlook the sparsity information during simulation and have limited backend support.  
As illustrated in~\figref{fig:motivation},
to leverage the sparsity and dynamic optimization opportunities,
our simulator proposes a novel sparsity-aware gate fusion technique and a dynamic kernel generation process that utilizes runtime matrix information for further optimizations.
Compared with prior quantum circuit simulators,
our tool enables more effective gate fusion with highly optimized kernels, significantly improving the simulation performance.
By integrating these optimizations and backend solutions, our proposed toolchain delivers cross-platform, scalable, and high-performance quantum simulations.

\section{Overview}\label{sec:overview}

\subsection{CAST Compilation}

\begin{figure*}[tb]
    \centering
    \includegraphics[width=0.95\linewidth]{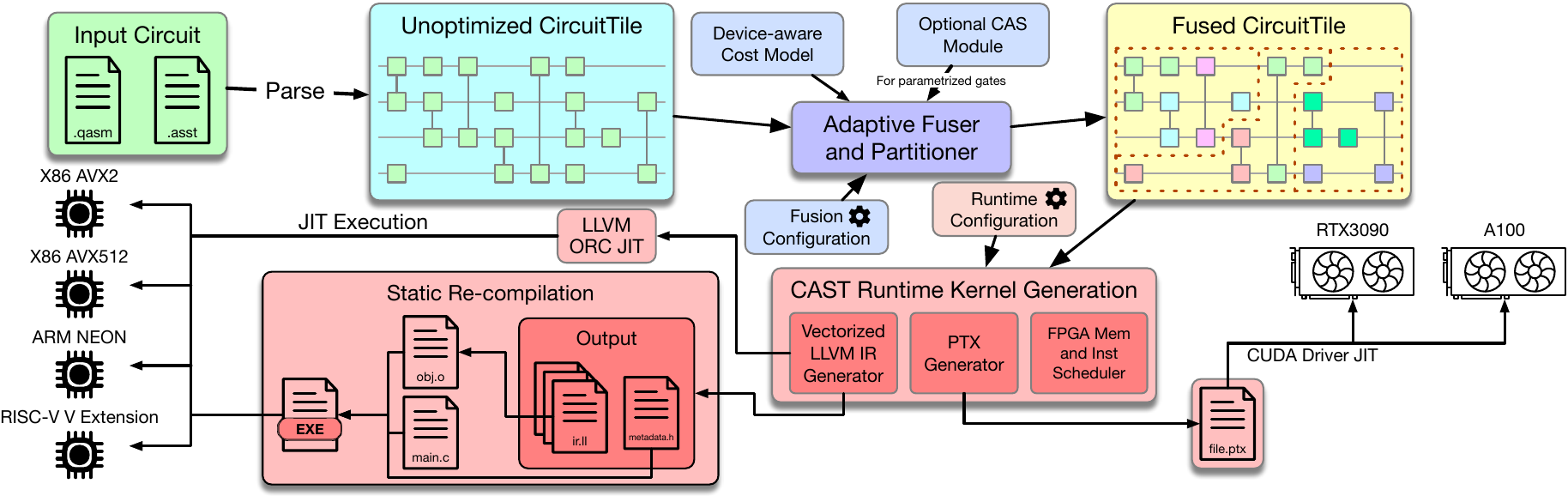}
    \caption{\CAST{} Workflow.}
    \label{fig:workflow}
\end{figure*}

\CAST{} compilation consists of two main phases: an adaptive gate fuser and partitioner that transpiles input circuits via a custom data structure \texttt{CircuitTile}, and a dynamic kernel generator that emits and optionally compiles hardware-specific instructions. 
A schematic diagram is given in~\figref{fig:workflow}.

\CAST{}'s workflow brings the following advantages: \begin{itemize}
    \item When targeting CPU based platforms, \CAST{} provides versatile SIMD capabilities via LLVM IR backend. This avoids the traditional ahead-of-time approach by hard-coding multiple vector intrinsics for different SIMD instruction sets and simulation precisions. 
    \item \CAST{} optimizes sparse gates during kernel generation. For example, all zero entries in the matrix are skipped in generated kernels. This brings a cross-platform solution for the previously infeasible task of preparing optimized kernels for fused gates.
    \item \CAST{} kernel generator is highly customizable and extendable. Kernel generators allow designers to obtain a consistent performance across platforms without repetitive work of implementing hardware-specific optimizations. Along the automation process, \CAST{} also provides a lot of room for exploration as shown in \tabref{tab:186}.
\end{itemize}
\vspace{-0.5em}



\begin{table}[hbt]
\renewcommand{\arraystretch}{1.0}
\renewcommand{\code}[1]{\fontsize{7}{6.5}\selectfont\texttt{#1}}
    \centering
    \caption{Part of \CAST{} Software Arguments.}
    \begin{tabularx}{\linewidth}{l|X}
    \textbf{Argument} & \textbf{Command line} \\\hline\hline
    Precision & \code{-f32, -f64} \\\hline
    \makecell[tl]{CPU\\vector size} & \makecell[tl]{\code{-S=<int>, -neon, -sse}\\\code{-avx2, -avx512}} \\\hline
    Gate fusion & \makecell[tl]{\code{-cost-model=<cost-model>, -fusion=<int>}\\\code{-zero-tolerance=<float>}\\\code{-one-tolerance=<float>}\\\code{-multi-traversal=<bool>}\\\code{-agglomerative=<bool>}} \\\hline
    \makecell[tl]{Kernel\\configurations} & \makecell[tl]{\code{-arch=<cpu/gpu>}\\\code{-use-fma, -use-fms, -use-pdep}\\\code{-alt-format, -sep-format}\\\code{-zero-tolerance-kernel=<float>}\\\code{-elem-sharing-threshold=<float>}\\\code{-use-imm-value=<bool>}\\\code{-use-const-mem-space=<bool>}\\\code{-force-dense-kernel=<bool>}} \\\hline
    \end{tabularx}
    \label{tab:186}
\end{table}

\section{Adaptive Sparsity-Aware Agglomerative Gate Fusion} \label{sec:gate_fusion}

\subsection{\code{CircuitTile} Data Structure}
Because the difficulty of gate fusion lies in preserving the commutation relation of quantum circuits, we design a custom data structure \code{CircuitTile} as an intermediate layer to represent quantum circuits and to perform gate fusion.

Internally, \code{CircuitTile} contains a linked list of arrays that we call \emph{tile}. Each array in the tile is called a \emph{row}. Every row is an array of null-able \code{GateBlock} pointers. A \code{GateBlock} is an aggregate of gates. Each gate keeps track of its left and right neighbors. 

To add a gate to the graph, we create a \code{GateBlock} consisting of solely this gate and append it to the tile. Tile spans vertically by convention, so to append a \code{GateBlock} to the tile, we insert it to the uppermost vacant row.


\subsection{Agglomerative Gate Fusion}

Gates with disjoint target qubit sets always commute with each other. So in the tile, blocks on the same row can be safely fused together. 
On the other hand, gates not on the same row can be fused together only when they are connected and on consecutive rows.

In other words, a pair of gate blocks can be fused in two scenarios. The first scenario is \textbf{Commuting Gate Fusion}, where the two blocks are on the same row. The second scenario is \textbf{Consecutive Gate Fusion}, where the two blocks are located in consecutive rows, and there is at least one direct connection spanning between them.

\CAST{} employs tile-based iteration to perform gate fusion. The pseudo-code is provided in Algorithm \ref{alg:1}. This algorithm takes in a CircuitTile $T$ and an unsigned integer $k$ as the input. It traverses through all gate blocks and fuse the blocks. During the execution, the number of qubits of the fused gate should stay below $k$ to avoid fragments. The algorithm returns \textbf{true} if and only if there is at least one pair of gate block merged during the execution.



\begin{algorithm}
\caption{Tile Traversal in Agglomerative Gate Fusion}
\fontsize{8.5}{9.0}\selectfont
\label{alg:1}
\begin{algorithmic}[1]
\Function{Traverse}{$\textrm{CircuitTile $T$, unsigned $k$}$}
    \State{$\delta \gets \textbf{false}$}
    \For{$r \in [0 .. N_\mathrm{rows}-1]$}
        \For {$q \in  [0 .. N_\mathrm{qubits}-1]$}
            \State $b_{\top} \leftarrow T_{r,q}$
            \If {$b_{\top} \neq \varnothing$}
                \State $b_{\bot} \leftarrow T_{r+1,q}$
                \If {$b_{\bot} \neq \varnothing$}
                    \State \Call{MoveBlock}{$b_{\top}, b_{\bot}$}
                \EndIf
                \If {\Call{Fusible}{$b_{\top}, b_{\bot}, k$}}
                    \State \Call{Fuse}{$b_{\top}$, $b_{\bot}$} \Comment{Consecutive Fusion}
                    \State $\delta \gets \textbf{true}$ 
                \EndIf
            \EndIf
        \EndFor
        \For {$q \in  [1 .. N_\mathrm{qubits}-1]$}
            \State $b_{\top} \leftarrow T_{r,q-1}$
            \State $b_{\bot} \leftarrow T_{r,q}$
            \If {\Call{Fusible}{$b_{\top}, b_{\bot}, k$}}
                \State \Call{Fuse}{$b_{\top}$, $b_{\bot}$} \Comment{Commuting Fusion}
                \State $\delta \gets \textbf{true}$ 
            \EndIf
        \EndFor
    \EndFor
    \State \Return{$\delta$}
\EndFunction
\end{algorithmic}
\end{algorithm}

\begin{figure}[t]
\begin{minipage}[b]{0.17\linewidth}
\includegraphics[width=\linewidth]{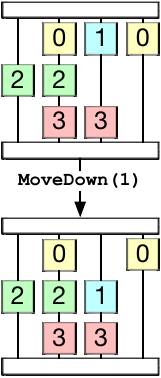}
\vspace{-1.5em}
\caption{\code{MoveDown}.}
\label{fig:537}
\end{minipage}
\hfil
\begin{minipage}[b]{0.77\linewidth}
\includegraphics[width=\linewidth]{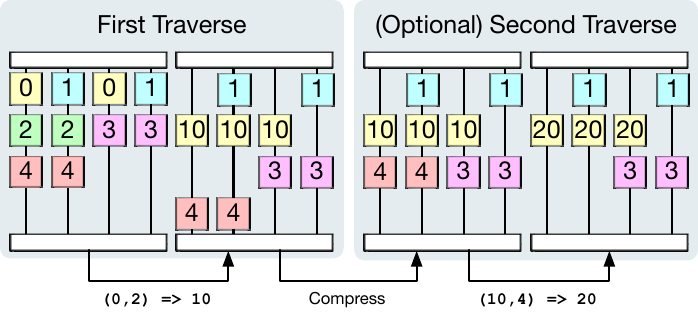}
\vspace{-1.5em}
\caption{Tile traversal.}
\label{fig:543}
\end{minipage}
\end{figure}

The algorithm actively relocates blocks to create fusion opportunities. Specifically, when the traversal reaches a gate block in the \(i\)-th row and the space below it in the \((i+1)\)-th row is unoccupied, the algorithm moves the gate block down to the \((i+1)\)-th row to fill the space. This relocation allows the block to potentially fuse with a block in the \((i+2)\)-th row. For example, \figref{fig:537} gives an illustration of such a case. In the compact tile above, blocks 1 and 3 are not in consecutive rows. Since the position immediately below block 1 is empty (line 8 in Algorithm \ref{alg:1}), block 1 can be moved down. Block 1 will then fuse with block 3 when we traverse to the new position of block 1.


To place fused blocks back into the tile,
the $\Call{Fuse}{}$ procedure first removes $b_{\top}$ and $b_{\bot}$. If the removal creates enough space in rows $r$ or $(r+1)$ to accommodate the fused block, the fused gate block occupies the corresponding space (prioritize on row $(r+1)$). Otherwise, the $\Call{Fuse}{}$ procedure inserts a new row between the two rows and places the fused block in the new row. Note that the creation of the new row can leave empty spaces (\figref{fig:543}). To maintain the compactness of the \code{CircuitTile}, we compress it at the end of each tile traversal.

Given a specified maximum size $k_{\max}$ of fused gates, the agglomerative scheme iteratively increases the allowed size of fused gates from 2 to $k_{\max}$.
We found that such an incrementing scheme helps reduce fragments (small gates remaining after the algorithm finishes) and leaves with overall sparser gates.

\begin{figure}
\includegraphics[width=0.9\linewidth]{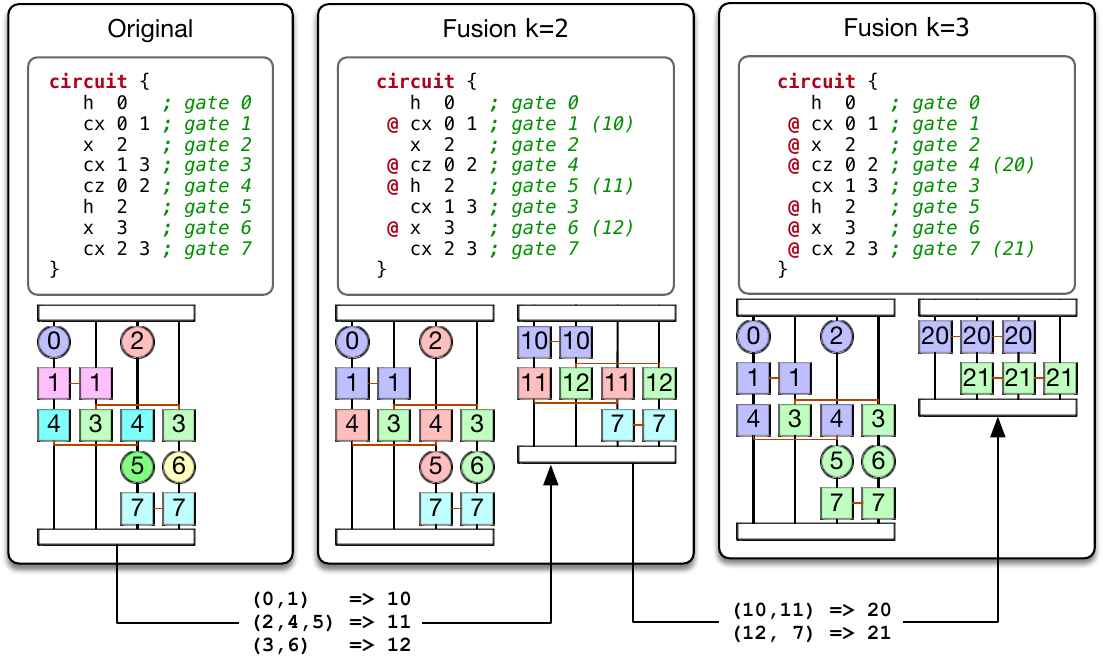}
\vspace{-0.5em}
\caption{Demonstration of agglomerative gate fusion. Each top block shows \CAST{} serialization of the \code{CircuitTile}. The \code{@} symbol denotes fusing gates together.}
\label{fig:504}
\end{figure}

\subsection{Adaptive Fusibility Check}
\label{sec:safc}



Adaptive gate fusion traverses the tile the same way as before, but adds extra criteria in fusion eligibility checks.
For CPU and GPU based adaptive gate fusion, we define a cost model upon the total operation count and the number of blocks present in the fused circuit.

The \emph{operation count} captures the sparsity of gates, and is defined as the number of run-time operations needed to perform matrix-vector multiplication. Take single-qubit gates as an example, the matrix-vector multiplication~\eqref{eq:330}
corresponds to updating four variables (registers) in each iteration.
For machines with support of both Fused Multiplication Addition (FMA) and Fused Multiplication Subtraction (FMS), every non-zero entry in the matrix requires $2$ operations in run-time. A dense $k$-qubit gate has $2^{2k+1}$ entries, and thus its operation count is $2^{2k+2}$. On the other hand, the operation count for sparse gates such as Pauli gates is far less than the worst case described above.

\begin{figure}[hbt]
    \centering
    \vspace{-1em}
    \includegraphics[width=0.8\linewidth]{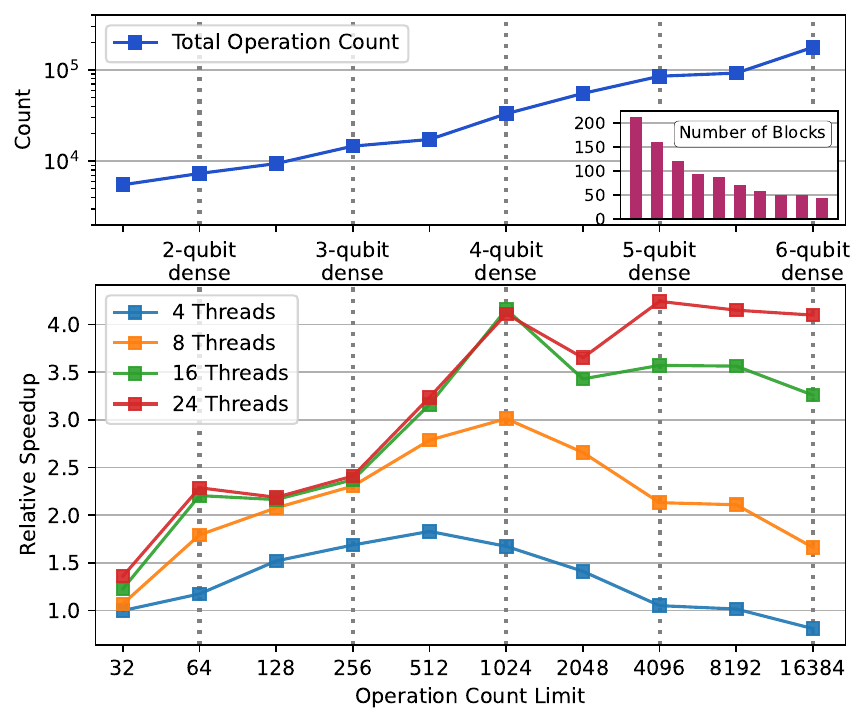}
    \vspace{-1em}
    \caption{Operation Count versus Simulation Time. Reference time is measured at the 4-thread 32-op-count-limit experiment. All experiments were carried out on CPU-based 30-qubit statevectors with double precision.}
    \label{fig:579}
\end{figure}

\figref{fig:579} shows one example. For a fixed number of threads, there is a peak performance point when we vary gate fusion aggressiveness. Left to the peak point is usually when memory bottlenecks speed, and right to the peak point is usually when computation power bottlenecks speed. 

\CAST{} provides several preset fusion configurations for CPU and GPU based platforms. Users are also encouraged to use the our provided \code{CostModel} class to conduct benchmarks for a cost model specialized to their hardware platform. \code{CostModel} will measure and record the running time of several quantum gates with varying sparsity and size and using different number of threads (for CPUs) and block dimensions (for Nvidia GPUs). These data will then by used in the fusion algorithm for adaptive fusibility checks. For example, on one of our CPU platforms, \code{CostModel} finds 6-qubit dense gates take significantly longer time to simulate than 5-qubit dense gates. So it decides to limit the operation count per gate to 4096 (equaling to that of a 5-qubit dense gate) while allowing fused gates to act on up to 7 qubits. As shown in~\tabref{tab:fusion_config}, in a sample Random Quantum Circuit (RQC), the adaptive configuration achieves a comparable compression ratio with a significantly reduced operation count.

\begin{table}[hbt]
    \centering
    \caption{Fusion Configuration Comparison.}
    \begin{tabular}{c|c|c|c}
    Configuration & \makecell{Compression\\Ratio} & \makecell{Fusion Time\\(ms)} & \makecell{Operation\\Count} \\\hline\hline
    Adaptive & 22.1x & 1423 & 1.05M \\
    Size-only & 23.2x & 988 & 1.74M \\
    Non-agglomerative & 23.2x & 6660 & 3.28M
    \end{tabular}
    \label{tab:fusion_config}
\end{table}

\section{Dynamic Kernel Generation and Compilation} \label{sec:kernel_gen}
\subsection{Automated Kernel Generation}

\CAST{} automates the kernel generation process for gates with various sizes and target qubits. Each \CAST{} CPU kernel takes 4 arguments and each GPU kernel takes 2 arguments. These arguments are summarized in the following list:
\begin{enumerate}
    \item Pointer to the statevector array.
    \item Pointer to the matrix array. This is to support simulating parametrized quantum circuits.
    \item Loop counter begin index, present only in CPU kernels.
    \item Loop counter end index, present only in CPU kernels.
\end{enumerate}

\CAST{} backend kernel generator starts by emitting LLVM IR.
In CPU kernels, the loop counter runs in $[0,2^{n-k-s})$, where $n$ is the number of qubits, $k$ is the gate size, and $S=2^s$ specifies vector size to be used in CPU-based SIMD instructions. This range can then be distributed to different threads for parallel execution. GPU kernels do not take loop counter as arguments because the loop counter is effectively tracked via built-in \code{blockIdx} and \code{threadIdx} variables. Each GPU thread has an effective vector size of 1. So the following IR generation process also works for GPUs with $s=0$.

For a $k$-qubit gate, we need to update $2^{k+1}$ vector amplitude registers in the main loop. 
The address and contiguity pattern of the amplitudes heavily depend on SIMD size and the set of target qubits.

\begin{figure}[hbt]
    \centering
    \vspace{-1em}
    \includegraphics[width=0.75\linewidth]{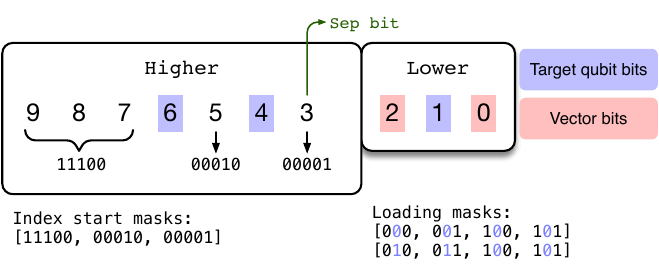}
    \vspace{-1em}
    \caption{Illustration when $s=2$ and target qubits are $\{1,4,6\}$.}
    \label{fig:322}
\end{figure}

Loading amplitude registers has two sub-steps: first calculate the start index from loop counter $t$, then depending on amplitude format, load $2^k$ real and imaginary amplitude registers each.
We first split target qubits into two sets, the higher set and the lower set. Start by enumerating indices from 0. Color target qubit indices blue and color the remaining $s$ smallest numbers red. Everything smaller than or equal to the largest red index is on the lower side (as shown in \figref{fig:322}).

Let $Q$ be the set of target qubits, $H$ be the set of all higher indices, and $L$ be the set of all lower indices. For some convenient numbers, let $k=|Q|$ be the size of the gate, and let $k_L=|Q\cap L|$ and $k_H=|Q\cap H|$ be the number of target qubits inside lower and higher set respectively. Then
\begin{itemize}
    \item If $k_L=0$, no runtime shuffling is needed both in loading and storing.
    \item Otherwise, the loop body loads (and stores) a total of $2^{k_H+1}$ length-$2^{k_L+s}$ LLVM vectors. These vectors will then be shuffled to $2^{k+1}$ length-$2^s$ vectors before matrix-vector multiplication. So on the LLVM IR level, when $k_L$ is non-zero, shuffled loading works by loading larger LLVM registers and let the assembler split these registers and assign hardware-native shuffling instructions.
    \item Depending on hardware support, at most $3k_H + 1$ operations are needed to compute the start index:
    \begin{equation}\label{eq:720}
        \code{startIdx} = \sum_{i=0}^{|H\cap Q|} \code{(t \& masks[i]) {<}{<} i},
    \end{equation}
    where \code{masks} is a compile-time known array\footnote{For machines that support BMI2 instruction set, a single runtime PDEP operation can replace Eq. \eqref{eq:720}. This is automatically configured in \CAST{} argument \code{-use-pdep}.}.
\end{itemize}


\subsection{JIT Execution and Re-compilation}

As part of design flexibilities, \CAST{} provides two modes for simulation on CPU-based platforms: JIT mode and static mode. JIT mode adopts LLVM ORC JIT engine to JIT-compile generated IR to native machine code for immediate execution. The static mode separate kernel generation and execution. It caches and saves generated IR files along with a header file of metadata. The saved IR files and metadata can then experience a compilation and linking phase to generate the final executable when needed.

GPU-based platforms follow a similar approach. The difference is that CUDA execution is internally JIT. \CAST{} provides the flexibility of choosing to load the generated PTX files into a CUDA context in runtime or saving it to the disk for future needs.

\section{Evaluation}\label{sec:eval}
\subsection{Setup}

\textbf{Software}\quad{}
In our evaluations we used the following software:


\begin{itemize}
    \item LLVM project: version 19.1.0. 
    \item Qiskit: version 1.2.4 with backend QiskitAer 0.15.1.
    \item Qibo: version 0.2.13; QiboJIT: version 0.1.7.
    \item Cirq: version 1.4.1; QSimCirq: version 0.21.0.
    \item Qulacs: version 0.6.11.
\end{itemize}


\textbf{Benchmark Circuits}\quad{}
We used six types of benchmark circuits in our evaluations.
\begin{itemize}
    \item Quantum Fourier Transform (QFT)~\cite{shor1994algorithms}.
    \item Alternating Layered Ansatz (ALA)~\cite{cerezo2021cost}.
    \item Random Quantum Circuit (RQC)~\cite{boixo2018characterizing}.
    \item Quantum Volume Circuit (QVC)~\cite{cross2019validating}.
    \item Instantaneous Quantum Polynomial (IQP)~\cite{bremner2017achieving}.
    \item Hamiltonian Evolution Simulation (HES)~\cite{low2017optimal}.
\end{itemize}

Among the six types of circuits, QFT, RQC, IQP, and HES are relatively more sparse. So we call them \emph{sparse-class} benchmarks. Correspondingly, ALA and QVC consist of mostly dense single-qubit gates, so we call them \emph{dense-class} benchmarks.

\subsection{Ablation Study on Adaptive Gate Fusion}

To evaluate the performance improvement brought by our adaptive fusion algorithm, we perform an ablation study on different benchmark circuits with three fusion configurations: no fusion, preset size-only fusion, and adaptive fusion. We record the running time of each experiment under each fusion configuration and normalize it by the running time taken under the no fusion configuration.

\begin{figure}[hbt]
    \centering
    \includegraphics[width=\linewidth]{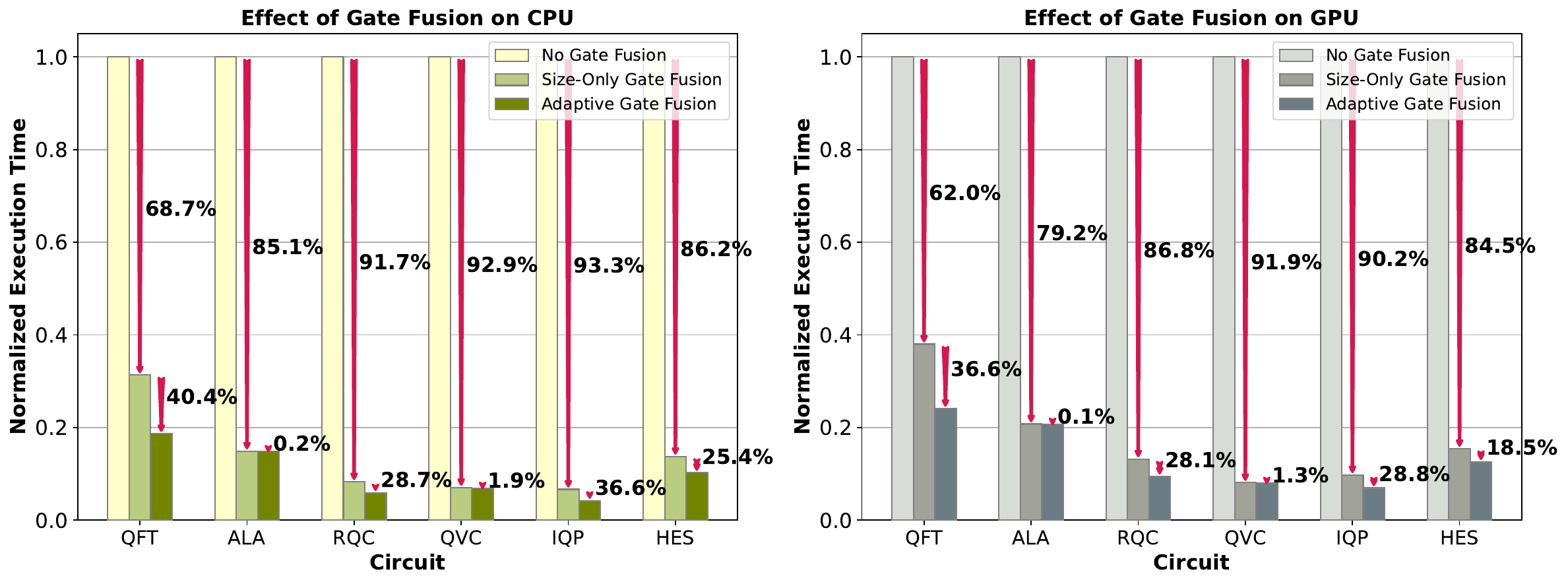}
    \vspace{-2em}
    \caption{Execution time of quantum simulation with different fusion strategies.}
    \vspace{-3mm}
    \label{fig:fusion_ablation}
\end{figure}

As shown in~\figref{fig:fusion_ablation}, compared with the original circuit, size-only fusion optimization can reduce the running time by 68.7\% $\thicksim$ 93.3\% (resp. 62.0\% $\thicksim$ 91.9\%) over different benchmarks on the CPU (resp. GPU) platform. Compared with traditional size-only fusions, our new adaptive fusion algorithm obtains comparable performance on dense-class benchmarks, and can further reduce the running time by 28.7\% $\thicksim$ 40.4\% (resp. 18.5\% $\thicksim$ 36.6\%) on sparse-class benchmarks on the CPU (resp. GPU) platform.

\begin{table*}[t]
\fontsize{7.2}{8.2}\selectfont
\addtolength{\tabcolsep}{-0.2em}
\renewcommand{\arraystretch}{1.15}
    \caption{Comparison of Single Gate Performance.}
    \begin{minipage}{\linewidth}
    \centering
    \begin{tabular}{|r|rrr|rrr||rrr|rrr||rrr|rrr|} \hline
    \multirow{3}{*}{\textbf{Gate}}
    & \multicolumn{6}{c||}{Apple M2 (NEON)}
    & \multicolumn{6}{c||}{Intel(R) Core(TM) i9-11900K (AVX512)} 
    & \multicolumn{6}{c|}{Nvidia RTX3090 (CUDA)}\\\cline{2-19}
   
    & \multicolumn{3}{c|}{\code{f32} Mode (GiBps)}
    & \multicolumn{3}{c||}{\code{f64} Mode (GiBps)}
    & \multicolumn{3}{c|}{\code{f32} Mode (GiBps)}
    & \multicolumn{3}{c||}{\code{f64} Mode (GiBps)}
    & \multicolumn{3}{c|}{\code{f32} Mode (GiBps)}
    & \multicolumn{3}{c|}{\code{f64} Mode (GiBps)} \\
    & Qiskit & Qibo & Ours & Qiskit & Qibo & Ours 
    & Qiskit & QSim & Ours & Qiskit & QSim & Ours 
    & Qiskit & Qulacs & Ours & Qiskit & Qulacs & Ours \\\hline

    $\gH^{\otimes 2}$
& 1.74 & 3.20 & \bf32.9 
& 3.46 & 6.04 & \bf34.1
& 1.28 & 15.6 & \bf17.3 
& 2.50 & N/A & \bf17.1
& 75.2 & N/A & \bf416 
& 181 & 150 & \bf417
\\
    $\gUthree^{\otimes 2}$
& 1.74 & 3.18 & \bf20.7
& 3.45 & 5.87 & \bf21.1
& 1.28 & 15.5 & \bf16.5 
& 2.62 & N/A & \bf16.2
& 75.4 & N/A & \bf418 
& 180 & 151 & \bf320
\\
    $\gH^{\otimes 3}$
& 0.666 & 3.21 & \bf22.0 
& 1.22 & 5.95 & \bf22.8
& 0.830 & 12.5 & \bf15.6 
& 1.56 & N/A & \bf14.9
& 18.9 & N/A & \bf396 
& 80.6 & 81.4 & \bf308
\\
    $\gUthree^{\otimes 3}$
& 0.663 & 3.11 & \bf8.38 
& 1.28 & 6.11 & \bf8.50
& 0.838 & \bf12.4 & 12.3 
& 1.56 & N/A & \bf12.7
& 18.9 & N/A & \bf393 
& 80.6 & 81.3 & \bf149
\\\hline
    \end{tabular}
    \end{minipage}
    \label{tab:single_gate}
\end{table*}

\subsection{Cross-platform Backend}
To demonstrate the effectiveness of \CAST{} kernel generators on various platforms and vairous precisions, we benchmark and compare the performance of dynamic kernels generated by \CAST{} and traditional ahead-of-time kernels as in Qiskit, Qibo, QSim, and Qulacs on the task of simulating gates with various sparsity on 28-qubit statevectors.

As reported in~\tabref{tab:single_gate}, our \CAST{} backend achieves comparable performance on dense two-qubit and three-qubit gates compared with QSimCirq on AVX-512 platform \code{f32} mode and outperforms Qiskit, Qibo, QSim, and Qulacs on other benchmarks. This demonstrates our dynamically generated \CAST{} backend has great versatility over different CPU and GPU platforms and simulation precisions. In addition, our backend generally runs faster on sparser gates, which verifies \CAST{} is able to automatically explore the sparsity pattern of gates.

\subsection{Compilation Overhead} \label{sec:comp_overhead}
Compared with traditional approaches, \CAST{} adopts extra code generation and re-compilation phases. We evaluate the impact of software overhead in this set of experiments.

\begin{figure}[hbt]
    \centering
    \includegraphics[width=0.99\linewidth]{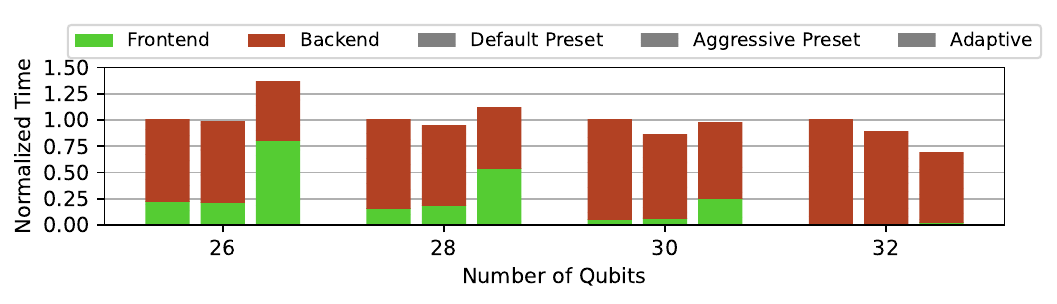}
    \vspace{-20pt}
    \caption{Front-end overhead.}
    \label{fig:CAST_overhead}
\end{figure}

We perform end-to-end simulations of RQCs with various number of qubits and record the normalized time spent in \CAST{} front-end and hardware backend respectively. The front-end consists of the full process from parsing input files to obtaining executable. Normalized time is calculated by dividing the total end-to-end time spent with default fusion preset by the number of gates in the original circuit. As reported in~\figref{fig:CAST_overhead}, the ratio of \CAST{} front-end overhead diminishes quickly as number of qubits increases. For example, in the 32-qubit experiment, the front-end of \CAST{} only takes 0.59\% (resp. 3.11\%) of the end-to-end running time under the default (resp. adaptive) fusion configuration. This demonstrates \CAST{} has great scalability in simulations. On the other hand, this set of experiments also suggests our adaptive fusion scheme exchanges slight front-end overhead for a great performance gain in running time.

\subsection{Full Circuit Simulation}
We benchmark our \CAST{} on full circuit simulations on CPU and GPU platforms.

\textbf{CPU Performance}\quad{}
We benchmark our simulator against Qiskit, Qibo, and QSimCirq on 32-qubit circuit simulations with multi-threading enabled. We record the execution time as reported by each simulator. Results are presented in~\tabref{tab:multithread}. 

For single-precision mode simulations, compared with QSimCirq, our simulator achieves comparable performance on dense-class circuits and up to 3.68x speedup on sparse-class circuits such as QFT-30. This indicates our new sparsity-aware gate fusion algorithm is particularly effective in sparse circuits while still attaining a top performance on dense circuits. Over different types of benchmarks, \CAST{} achieves 3.58x $\thicksim$ 8.03x (resp. 2.14x $\thicksim$ 4.92x) speedup than Qiskit and 3.60x $\thicksim$ 17.4x (resp. 2.29x $\thicksim$ 12.0x) speedup than Qibo in single-precision (resp. double-precision) mode simulations.

\textbf{GPU Performance}\quad{}
We benchmark our simulator against Qiskit and Qulacs on 30-qubit circuit simulations. We record the execution time reported in each software. Qulacs has its own CUDA backend that only supports double-precision simulations. The GPU backend of Qiskit is cuQuantum~\cite{faj2023quantum}, which is developed by Nvidia and supports both \code{f32} and \code{f64} format simulation (\tabref{tab:149}). 

As shown in~\tabref{tab:gpu}, over different types of benchmarks, \CAST{} achieves 6.71x $\thicksim$ 39.3x speedup compared with Qiskit in single-precision mode simulations, and 2.08x $\thicksim$ 10.3x speedup compared with Qiskit and Qulacs in double-precision mode simulations.

The results demonstrate our toolchain has great performance and versatility in both single and double-precision modes and on different hardware platforms.

\begin{table}[tb]
\fontsize{7.5}{8.5}\selectfont
\addtolength{\tabcolsep}{-0.15em}
\renewcommand{\arraystretch}{1.15}
    \caption{Comparison of CPU Performance.}
    \begin{minipage}{\linewidth}
    \centering
    \begin{tabular}{|r|rrrr|rrrr|} \hline
      \multirow{3}{*}{\textbf{Circuit}} & \multicolumn{8}{c|}{AMD EPYC 7543 (32-core, AVX2)} \\\cline{2-9}
      & \multicolumn{4}{c|}{Single Precision (s)} & \multicolumn{4}{c|}{Double Precision (s)} \\
      & Qiskit & Qibo & QSim & Ours & Qiskit & Qibo & QSim & Ours \\\hline

    \bf QFT-32
    & 145 & 427 & 98.8 & \bf27.4 
    & 190 & 574 & N/A & \bf51.4\\ 
    \bf ALA-32
    & 876 & 1011 & 204 & \bf176  
    & 956 & 1342 & N/A & \bf337\\ 
    \bf RQC-32
    & 190 & 237 & 71.1 & \bf26.8  
    & 205 & 304 & N/A & \bf55.8 \\ 
    \bf QVC-32
    & 409 & 430 & \bf110 & 113  
    & 467 & 572 & N/A & \bf223 \\ 
    \bf IQP-32
    & 589 & 327 & 82.1 & \bf25.3  
    & 1087 & 420 & N/A & \bf49.6 \\ 
    \bf HES-32
    & 1918 & 1867 & 605 & \bf201  
    & 3457 & 2176 & N/A & \bf408 \\ 
    \hline
    \end{tabular}
    \end{minipage}
    \label{tab:multithread}
\end{table}

\begin{table}[tb]
\fontsize{7.5}{8.5}\selectfont
\renewcommand{\arraystretch}{1.15}
    \caption{Comparison of GPU Performance.}
    \begin{minipage}{\linewidth}
    \centering
    \begin{tabular}{|r|rrr|rrr|} \hline
      \multirow{3}{*}{\textbf{Circuit}} & \multicolumn{6}{c|}{Nvidia RTX3090} \\\cline{2-7}
      & \multicolumn{3}{c|}{Single Precision (s)}
      & \multicolumn{3}{c|}{Double Precision (s)} \\
      & Qiskit & Qulacs & Ours & Qiskit & Qulacs & Ours \\\hline

    \bf QFT-30
    & 16.4 & N/A & \bf2.44   
    & 12.0 & 52.4 & \bf5.44 \\ 
    \bf ALA-30
    & 131 & N/A & \bf9.83   
    & 99.2 & 96.6 & \bf28.3\\ 
    \bf RQC-30
    & 307 & N/A & \bf15.9   
    & 220 & 283 & \bf79.7 \\ 
    \bf QVC-30
    & 198 & N/A & \bf5.04   
    & 161 & 53.1 & \bf25.5 \\ 
    \bf IQP-30
    & 41.0 & N/A & \bf2.01   
    & 45.2 & 52.4 & \bf5.08 \\ 
    \bf HES-30
    & 683 & N/A & \bf81.4   
    & 479 & 1260 & \bf173 \\ 
    \hline
    \end{tabular}
    \end{minipage}
    \label{tab:gpu}
\end{table}

\section{Conclusion}
This paper proposes a high-performance cross-platform quantum circuit simulation framework \CAST{} (Cross-platform Adaptive Schr\"odinger-style Simulation Toolchain).
\CAST{} automates circuit optimization and kernel generation process by introducing a novel adaptive gate fusion and optimization algorithm and cross-platform backend kernel generators. Extensive evaluations suggest \CAST{} can achieve higher performance over the leading industrial and academic quantum circuit simulators on large-size quantum circuit simulations with a minimized compilation overhead.

\section{Acknowledgment}
The support of UK EPSRC (grant number EP/W032635/1, 
EP/V028251/1, EP/S030069/1 and EP/X036006/1), Intel and
AMD is gratefully acknowledged.

\vfill
\newpage
\clearpage
\bibliographystyle{IEEEtran}
\bibliography{dac25}

\end{document}